\documentclass[aps,prl,twocolumn,showpacs,preprintnumbers,superscriptaddress]{revtex4}%
\usepackage{amssymb}
\usepackage{graphicx}
\usepackage{dcolumn}
\usepackage{bm}
\usepackage{amsmath}
\usepackage{amsfonts}%
\begin{document}
\title{Observation of Hysteretic Transport Due to Dynamic Nuclear Spin Polarization\\
in a GaAs Lateral Double Quantum Dot}
\author{Takashi Kobayashi}
\affiliation{NTT\,Basic\,Research\,Laboratories,\,NTT\,Corporation,\,3-1\,Morinosato-Wakamiya,\,Atsugi\,243-0198,\,Japan}
\affiliation{Department of Physics, Tohoku University, Sendai 980-8578, Japan}
\author{Kenichi Hitachi}
\affiliation{NTT\,Basic\,Research\,Laboratories,\,NTT\,Corporation,\,3-1\,Morinosato-Wakamiya,\,Atsugi\,243-0198,\,Japan}
\author{Satoshi Sasaki}
\affiliation{NTT\,Basic\,Research\,Laboratories,\,NTT\,Corporation,\,3-1\,Morinosato-Wakamiya,\,Atsugi\,243-0198,\,Japan}
\affiliation{Department of Physics, Tohoku University, Sendai 980-8578, Japan}
\author{Koji Muraki}
\affiliation{NTT\,Basic\,Research\,Laboratories,\,NTT\,Corporation,\,3-1\,Morinosato-Wakamiya,\,Atsugi\,243-0198,\,Japan}
\date{\today}

\begin{abstract}
We report a new transport feature in a $\text{GaAs}$ lateral double quantum dot that emerges for magnetic-field sweeps and shows hysteresis due to dynamic nuclear spin polarization (DNP). 
This DNP signal appears in the Coulomb blockade regime by virtue of the finite inter-dot tunnel coupling and originates from the crossing between ground levels of the spin triplet and singlet extensively used for nuclear spin manipulations in pulsed gate experiments. 
The magnetic-field dependence of the current level is suggestive of unbalanced DNP between the two dots, which opens up the possibility of controlling electron and nuclear spin states via DC transport.
\end{abstract}

\maketitle





Electron spin dynamics in a semiconductor quantum dot (QD) is affected by the nuclear spin (NS) ensemble of the host material through the effective magnetic field (Overhauser field) produced via the hyperfine (HF) interaction \cite{Petta, KoppensQ, Nowack, Coish, Khaetskii}. This is important in view of quantum information processing using physical qubits based on electron spins in QDs \cite{Petta, KoppensQ, Nowack}. In particular, when a qubit is encoded in the $S_{z}=0$ subspace of singlet and triplet states in a double QD (DQD), even a small difference in the Overhauser field $\Delta\vec{B}_{N}\equiv\vec{B}_{N,L}-\vec{B}_{N,R}$ between the left and right dots can drastically affect the time evolution of the qubit, with its coherence time limited by the fluctuation in $\Delta\vec{B}_{N}$. On the other hand, electron-nuclear spin coupled dynamics that takes place around a spin triplet-singlet ($T$-$S$) crossing has been exploited as a resource for preparing both electron and nuclear spin states in a controlled manner \cite{Petta2, Foletti, Reilly, Klauser, Ribeiro}. Using various pulsed gate sequences, $|\Delta B_{N}^{z}|$ has been enhanced or its fluctuation squeezed, which respectively allow for rapid and universal control \cite{Foletti} or a longer coherence time \cite{Reilly, Klauser,Ribeiro} of the qubit. Such ambivalent roles of NSs have motivated numerous studies aimed at understanding and controlling NS dynamics in DQDs \cite{Petta2, Reilly, Klauser, Ribeiro, Vink, Foletti, Danon}.


DC transport in the spin blockade (SB) regime \cite{Ono2}, where the charge transport is governed by transitions between different spin states, is one of the powerful means for investigating the interplay between electron and nuclear spins in DQDs. Experiments have revealed current fluctuations arising from random dynamics of NSs \cite{Koppens, Churchill} and/or hysteresis due to dynamic nuclear polarization (DNP) \cite{Koppens, Churchill, Ono, Pfund}. However, since DC transport relies on sequential tunnel processes involving different charge states, such as $(m,n)\rightarrow(m+1,n-1)\rightarrow(m,n-1)\rightarrow(m,n)$ [where $(m,n)$ denotes the occupation of the left and right dots], previous studies have been mainly focused on examining cases where all relevant electronic states are within the transport window. Consequently, electron-nuclear spin dynamics at the $T$-$S$ crossing, which has been extensively used in pulsed gate experiments \cite{Petta2, Reilly, Foletti}, has been poorly explored in DC regime.


In this Letter, we report a new transport feature in the Coulomb blockade (CB) regime of a few-electron $\text{GaAs}$ lateral DQD. 
We observe a sudden increase in the dot current $I_{\text{dot}}$ in the CB regime near the $(1,1)$-$(2,0)$ charge boundary. 
The feature shows hysteresis indicative of dot state locking due to DNP associated with the crossing between the ground levels of the triplet and singlet. 
As the inter-dot tunnel coupling is reduced, the feature weakens and eventually disappears, which indicates that the hybridization between the $(1,1)$ and $(2,0)$ singlets is essential for the lifting of the CB. 
Magnetic-field dependence of $I_{\text{dot}}$ suggests that the DNP generates strong imbalance of NS polarization between the two dots, which our model shows to be possible via positive feedback mechanism.
Our results provide new insights into electron-nuclear spin dynamics in a DQD and open up the possibility of controlling electron and nuclear spin states via DC transport.


As shown in Fig.\,1(a), the DQD used in this study was defined with $\text{Ti/Au}$ gates in a GaAs/Al$_{0.3}$Ga$_{0.7}$As heterostructure containing a two-dimensional electron gas (density $2.2\times10^{15}$\,m$^{-2}$ and mobility $200$\,m$^{2}/$Vs) $80\ \text{nm}$ below the wafer surface. We applied a fixed source-drain bias $V_{\text{SD}}=-800\,\mu\text{V}$ and measured $I_{\text{dot}}$ as a function of gate voltages $V_{\text{PL}}$ and $V_{\text{PR}}$ and in-plane magnetic field $B_{\parallel}$ applied in the direction shown by the solid arrow in the figure. The energy levels in the two dots were controlled with $V_{\text{PL}}$ and $V_{\text{PR}}$, while the inter-dot tunnel coupling $t$ was tuned with $V_{\text{C}}$. The actual values of $(m,n)$ were determined using a side-coupled quantum point contact charge sensor. All measurements were carried out with the sample mounted in a dilution refrigerator with a base temperature of $40\ \text{mK}$.


We first show SB characteristics of our DQD at $B_{\parallel}=0$. Figure\,1(b) shows a spectrum of $I_{\text{dot}}$ measured around the $(1,1)$-$(2,0)$ charge boundary as a function of $V_{\text{PL}}$ and $V_{\text{PR}}$. $I_{\text{dot}}$ is clearly suppressed inside the trapezoid region at the bottom of the overlapping triangular regions (bias triangles), in which electronic transport is supposed to be free of CB. This suppression of $I_{\text{dot}}$ is the manifestation of SB \cite{Ono2, Koppens}, which originates from the energy and spin conservation relevant to the two-electron states shown in Figs.\,1(c) and 1(d) as a function of the energy detuning $\varepsilon$ between the two dots measured from the $(2,0)$-$(1,1)$ charge degeneracy point [white arrow in Fig.\,1(b)]. For the $(2,0)$ charge configuration, the Pauli exclusion principle makes the triplet [$(2,0)T$] much higher in energy than the singlet [$(2,0)S$] with an energy gap $\Delta _{ST}\approx350\,\mu\text{eV}$, making $(2,0)T$ energetically inaccessible from all other states for $\varepsilon<\Delta_{ST}$.  For the $(1,1)$ charge configuration, both singlet [$(1,1)S$] and triplets ($T_{i}$) are within the transport window [$i$ ($=0,\pm$) denotes the $z$-component (along $B_{\parallel}$) of the total spin]. 
Since the $T_{i}\rightarrow(2,0)S$ and $T_{i}\rightarrow(1,1)S$ transitions are prohibited by spin conservation, the sequential tunnel process $(1,1)\rightarrow(2,0)\rightarrow(1,0)\rightarrow(1,1)$ \cite{sequential} for $V_{\text{SD}}<0$ as used here is blocked once  one of the $T_{i}$'s is occupied.


\begin{figure}
[ptb]
\begin{center}
\includegraphics[width=\linewidth]{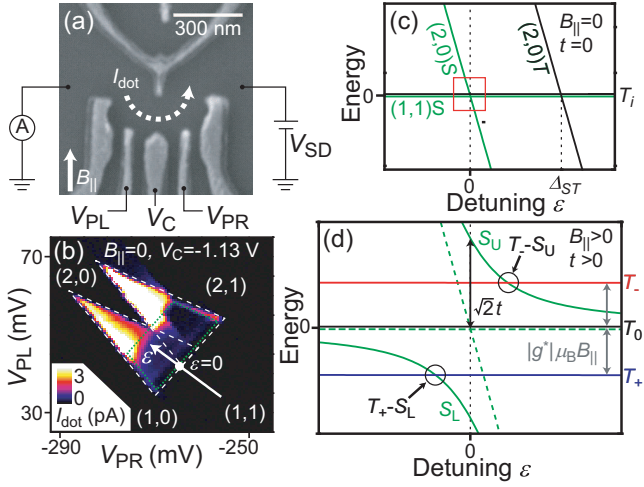}
\caption{(color online) (a) SEM image of the DQD sample. (b) $I_{\mathrm{dot}}$ spectrum as a function of $V_{\mathrm{PR}}$ and $V_{\mathrm{PL}}$ around the SB regime at $B_{\parallel}=0$ and $V_{\mathrm{C}}=-1.13$\,V. (c) Energies of relevant states around the SB regime at $B_{\parallel}=0$ and $t=0$. (d) Energy levels around $\varepsilon=0$ as shown by red-boxed region in (c) for $B_{\parallel}\neq0$ and $t\neq0$.}%
\end{center}
\end{figure}


The weak features seen in the SB regime, in turn, indicate leak current that occurs only via $T\rightarrow S$ spin transition. In GaAs DQDs, possible mechanisms for $T\rightarrow S$ transition include NS-mediated processes \cite{Erlingsson, Jouravlev} and ``NS-free" processes such as cotunneling \cite{Vorontsov} and spin-orbit coupling \cite{Khaetskii2, Danon2}. Contributions of these mechanisms can be distinguished by mapping $I_{\text{dot}}$ as a function of $\varepsilon$ and $B_{\parallel}$ \cite{Koppens, Pfund, Churchill}. The data shown in Fig.\,2(a) were obtained by sweeping $\varepsilon$ at each fixed value of $B_{\parallel}$. Consistent with the previous report for a $\text{GaAs}$ lateral DQD \cite{Koppens}, leak current appears only at $\varepsilon\approx0$ (along the $B_{\parallel}$ axis) or at $B_{\parallel}\approx0$ (along the $\varepsilon$ axis, but only at $\varepsilon>0$). As detailed in Refs.\,\cite{Koppens, Jouravlev}, both of these features originate from NS-mediated processes that take place at $T$-$S$ crossings; the former is presumably related to the $T_{-}$-$S_{U}$ and $T_{+}$-$S_{L}$ double crossings that occur only at $\varepsilon\approx0$, while the latter involves $S_{U}$ and all $T_{i}$'s ($i=0,\pm$), which become nearly degenerate at $B_{\parallel}\approx0$ and $\varepsilon>0$ [see Figs.\,1(c) and 1(d)]. Here, $S_{U(L)}$ denotes the upper (lower) branch of singlet eigenstates for finite $t$ hybridizing $(2,0)S$ and $(1,1)S$. Except for these regions with NS-mediated transport, $I_{\text{dot}}$ is small (less than $150\,\text{fA}$) at least at $\varepsilon<100\,\mu\text{eV}$, indicating that contributions of NS-free processes are minor here.


\begin{figure}
[ptb]
\begin{center}
\includegraphics[width=\linewidth]{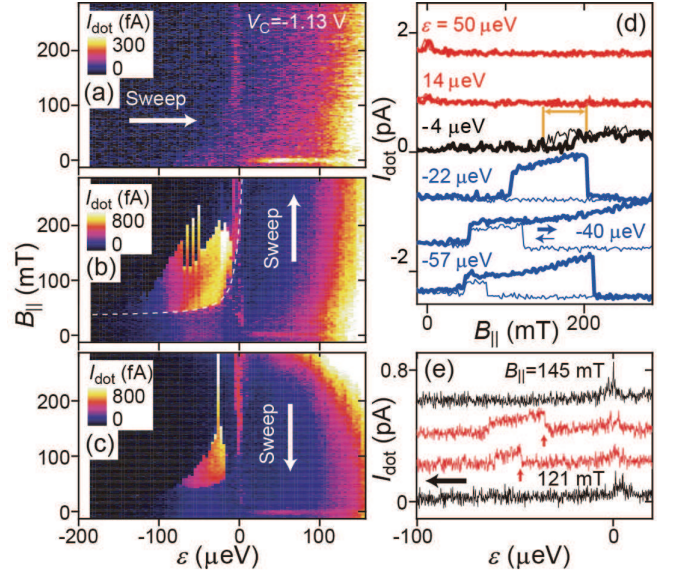}
\caption{(color online) $\varepsilon$ and $B_{\parallel}$ dependence of $I_{\mathrm{dot}}$ around the SB regime at $V_{\mathrm{C}}=-1.13$ V. (a) is obtained by $\varepsilon$ sweeps at each value of $B_{\parallel}$. (b) and (c) are obtained by up and down sweeps of $B_{\parallel}$ at each value of $\varepsilon$, respectively. (d) $I_{\mathrm{dot}}$ obtained by round sweeps of $B_{\parallel}$ around the SB regime at $V_{\mathrm{C}}=-1.13$\,V. Thick (thin) curves show the data for up (down) sweeps of $B_{\parallel}$. Curves are offset vertically by 0.8\,pA for clarity. (e) $I_{\mathrm{dot}}$ obtained by very slow $\varepsilon$ sweeps for several values of $B_{\parallel}$. Curves are offset vertically by 0.2\,pA for clarity.}%
\end{center}
\end{figure}



We find that strikingly different behavior shows up when $B_{\parallel}$, instead of $\varepsilon$, is swept at each fixed value of $\varepsilon$. When $B_{\parallel}$ is swept up [Fig.\,2(b)], a pronounced feature manifested as an enhancement of $I_{\text{dot}}$ emerges at $\varepsilon<0$ over a wide range of $\varepsilon$ and $B_{\parallel}$. It is noteworthy that the relevant region $\varepsilon<0$ is outside the bias triangles, which indicates that this new feature is emerging in the CB regime. Another important characteristic of this feature is its strongly hysteretic behavior with respect to the $B_{\parallel}$ sweep direction. As shown in Fig.\,2(c), the area of enhanced $I_{\text{dot}}$ becomes much smaller when $B_{\parallel}$ is swept down. Figure\,2(d) compares $I_{\text{dot}}$ traces for up and down sweeps at several values of $\varepsilon$. For up sweeps, the feature appears as a step-like enhancement of $I_{\text{dot}}$, which is then dragged over a wide magnetic-field range of $100\,\text{mT}$ or more. For down sweeps, such dragging is much weaker, or even the current enhancement itself can be absent (see the trace for $\varepsilon=-22\,\mu\text{eV}$).



We emphasize that the behavior of the hysteretic transport we observe at $\varepsilon<0$ is distinct from that at $\varepsilon\approx0$.
The trace for $\varepsilon\approx0$ ($=-4\ \mu\text{eV}$) in Fig.\,2(d) shows that drag is observed for both up and down sweeps. 
As reported for various material systems including not only $\text{GaAs}$ \cite{Koppens} but also for $^{13}\text{C}$-enriched carbon nanotubes \cite{Churchill}, such behavior occurs only near $\varepsilon \approx0$ and is presumably specific to the double crossings of $T_{+}$-$S_{L}$ and $T_{-}$-$S_{U}$, which occur only at $\varepsilon\approx 0$.


The observed hysteretic behavior suggests DNP induced by electron-nuclear spin flip flop \cite{Ono}. 
The drag of $I_{\text{dot}}$ enhancement indicates that, as a result of DNP, the system is locked into a state that triggers the onset of $I_{\text{dot}}$ while $B_{\parallel}$ is swept. 
Such locking is possible only when the average Overhauser field $B_{N}^{z}\equiv (B_{N,L}^{z}+B_{N,R}^{z})/2$ grows at a rate sufficient to cancel the variation of $B_{\parallel}$, a condition given by $|\dot{B}_{\parallel}|\ll|\dot{B}_{N}^{z}|$ and $\text{sgn}(\dot{B}_{\parallel})=-\text{sgn}(\dot{B}_{N}^{z})$, where the overdot denotes the time derivative. 
Hence, $\dot{B}_{N}^{z}<0$ must hold for up sweeps ($\dot{B}_{\parallel}>0$), which implies DNP into spin-up because of the negative $g$ factor ($g^{\ast}=-0.44$) and a positive HF coupling constant in $\text{GaAs}$ \cite{Paget}. 
The fact that the relevant feature appears at $\varepsilon<0$ (and at small $B_{\parallel}$) suggests the $T_{+}$-$S_{L}$ crossing as the cause of the DNP [Fig.\,1(d)]. 
An electron spin flip from up to down at the $T_{+}\rightarrow S_{L}$ transition would flop a NS from down to up, consistent with the required sign of DNP. 
Indeed, for up sweeps the position of the onset of $I_{\text{dot}}$ behaves similarly to the calculated position of the $T_{+}$-$S_{L}$ crossing as a function of $\varepsilon$ and $B_{\parallel}$ [dashed line in Fig.\,2(b)], given by $2|g^{\ast}|\mu_{B}(B_{\parallel}-B_{0})=\sqrt{8t^{2}+\varepsilon^{2}}+\varepsilon$, where $\mu_{B}$ is the Bohr magneton and $B_{0}$ is an adjustable parameter.



As shown in Fig.\,2(a), DNP is absent for $\varepsilon$ sweeps at typical sweep rates, irrespective of sweep direction. 
This is in part due to the energy resolution of $\varepsilon$ sweeps, which is generally much coarser than the energy scale relevant to $T$-$S$ mixing set by the statistical fluctuations of the Overhauser field $||\Delta\vec{B}_N||$ (typically $\sim 5\,$mT in a GaAs lateral DQD \cite{Koppens}). 
Indeed, for very slow and fine sweeps, we observe a similar $I_{\text{dot}}$ enhancement dragged by down sweeps, consistent with DNP into NS up [Fig.\,2(e)].
However, the DNP signal appears only in a narrow region and with much lower current level than for $B_{\parallel}$ sweeps. 
We believe that this is because changes in the electronic state induced by gate sweeps are not completely canceled by the Overhauser field, unlike in the case of $B_{\parallel}$ sweeps.


We note that the $T_{+}$-$S_{L}$ crossing has previously been identified in pulsed gate experiments combined with charge detection, where its position as a function of $\varepsilon$ and $B_{\parallel}$ has been mapped out \cite{Petta, Petta2, Foletti, Reilly}. 
Furthermore, passage through the $T_{+}$-$S_{L}$ crossing in a pulsed gate sequence has been harnessed to dynamically polarize nuclei \cite{Petta2, Foletti} or to prepare a preferred NS state \cite{Reilly}. 
In DC transport, on the other hand, due to CB at $\varepsilon <0$, which prohibits the charge transfer from $(1,1)$ to $(2,0)$ involved in the sequential tunnel process, the $T_{+}$-$S_{L}$ crossing is not usually observed \cite{T+SL}.
We suggest that in the present case CB is lifted by making the transition from $(1,1)$ to $(1,0)$ through $S_{L}$, that is, without making a real transition to $(2,0)S$. 
The transition from $S_{L}$ to $(1,0)$ is made possible by a finite $(2,0)S$ component in $S_{L}$ that exists even at $\varepsilon<0$ for finite $t$. 
The amplitude of this $(2,0)S$ component is given by $c(t,\varepsilon )\equiv|\langle(2,0)S|S_{L}\rangle|^{2}=[1+(\varepsilon/t)/\sqrt{(\varepsilon/t)^{2}+8}]/2$ \cite{Supplement}, indicating that it decreases with increasing $|\varepsilon|$ at $\varepsilon<0$. 
This qualitatively explains the observation that the DNP signal decays with increasing $|\varepsilon|$ [Fig.\,2(b)].


\begin{figure}
[ptb]
\begin{center}
\includegraphics[width=\linewidth]{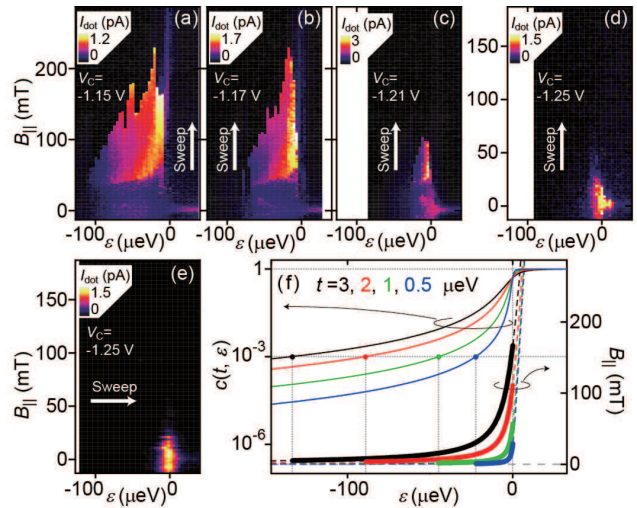}
\caption{(color online) (a)-(d) $I_{\mathrm{dot}}$ obtained by up sweeps of $B_{\parallel}$ at each value of $\varepsilon$: (a) $V_{\mathrm{C}}=-1.15$, (b) $-1.17$, (c) $-1.21$, and (d) -1.25\,V. (e) $I_{\mathrm{dot}}$ obtained by $\varepsilon$ sweeps at each value of $B_{\parallel}$ for $V_{\mathrm{C}}=-1.25$\,V. (f) Thin-solid curves show $\varepsilon$ dependence of $c(t,\varepsilon)=|\left\langle (2,0)S|S_{L}\right\rangle|^{2}$ for several values of $t$. Thin-broken curves show $\varepsilon$ dependence of $B_{\parallel}$ at the $T_{+}$-$S_{L}$ crossing point for each $t$. Thick-solid curves show the regions with $c(t,\varepsilon)\geq 10^{-3}$ at $\varepsilon <0$.}%
\end{center}
\end{figure}

The above equation shows that $c(t,\varepsilon)$ is a function of $\varepsilon/t$, suggesting that the ratio $\varepsilon/t$, rather than $\varepsilon$ itself, is the key parameter dictating the occurrence of DNP. Figures\,3(a)-3(d) show $I_{\text{dot}}$ obtained by up sweeps of $B_{\parallel}$ for several $V_{C}$ and thus for different values of $t$. While a DNP signal is present for $V_{C}\geq-1.21\,\text{V}$, as $V_{C}$ is made progressively more negative, the area of $\varepsilon$ where DNP is observed shrinks. By $V_{C}=-1.25\,\text{V}$, DNP is no longer discernible, where the overall feature becomes nearly identical to that for the $\varepsilon$ sweeps [Fig.\,3(e)]. The thin solid lines in Fig.\,3(f) depict $c(t,\varepsilon)$ as a function of $\varepsilon$ for several values of $t$. Since smaller $t$ gives smaller $c(t,\varepsilon)$ at $\varepsilon<0$, the range of $\varepsilon$ with a certain value of $c(t,\varepsilon)$ shrinks with decreasing $t$. Figure\,3(f) also shows the $T_{+}$-$S_{L}$ crossing point in the plane of $\varepsilon$ and $B_{\parallel}$ for several values of $t$, where regions with $c(t,\varepsilon)\geq 10^{-3}$ are highlighted with thick lines. These regions become shorter with decreasing $t$, in qualitative agreement with the observed $\varepsilon$ and $B_{\parallel}$ dependence of the onset of DNP [Figs.\,2(b) and 3(a)-3(c)].



It is noteworthy that, different from the previous report on a vertical DQD \cite{Ono}, DNP due to the $T_{-}$-$S_{U}$ crossing at $\varepsilon>0$ is not observed here. Note that DNP via the $T_{-}$-$S_{U}$ crossing is predominantly limited by the transition rate from $T_{+}$ to other states, which are higher in energy at a finite $B_{\parallel}$. Since this requires phonon absorption for energy conservation, it is strongly suppressed at low temperature in the absence of NS-free processes such as cotunneling, as is the case here. This contrasts with the case of the $T_{+}$-$S_{L}$ crossing, where DNP is mainly limited by the relaxation rate from $T_{-}$ to lower lying states via HF coupling, which can take place with phonon emission even at low temperature.


\begin{figure}
[ptb]
\begin{center}
\includegraphics[width=\linewidth]{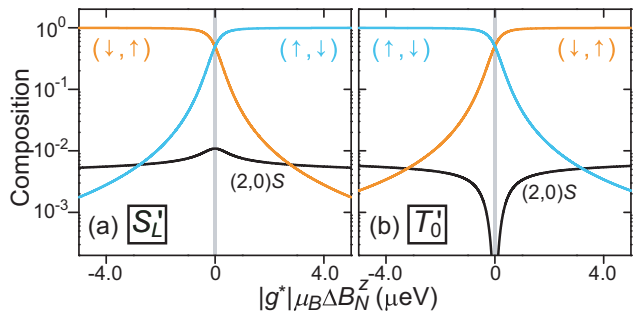}
\caption{(color online) Compositions of $(2,0)S$, $(\downarrow,\uparrow)$, and $(\uparrow,\downarrow)$ states in (a) $S_L^\prime$  and (b) $T_0^\prime$ as functions of $|g^{\ast}|\mu_B\Delta B_{N}^{z}$. $\varepsilon=-40\,\mu\text{eV}$ and $t=3\,\mu\text{eV}$ are assumed. Grey regions indicate the range of statistical fluctuations, $|\Delta B_{N}^{z}|\leq ||\Delta\vec{B}_N||/2$.}%
\end{center}
\end{figure}



While the occurrence of DNP and its dependence on $\varepsilon$ and $t$ can be explained as above, the dependence of $I_{\text{dot}}$ on $B_{\parallel}$ provides a deeper insight into the underlying transport mechanism. 
That is, when the dot state is locked via DNP, $I_{\text{dot}}$ does not remain constant, but increases with $B_{\parallel}$ [Fig.\,2(d)], which points to the existence of an additional degree of freedom that governs the transport. 
We propose a model that DNP generates imbalance $\Delta B_N^z$ in the NS polarization between the two dots and this finite $\Delta B_N^z$ enhances $I_{\text{dot}}$. 
When $\Delta B_N^z\neq 0$, $S_L$, $S_U$, and $T_0$ are no longer eigenstates of the system, and mix to form new eigenstates $S_L^\prime$, $S_U^\prime$, and $T_0^\prime$. 
Figures\,4(a) and 4(b) depict how the characters of $S_L^\prime$ and $T_0^\prime$ evolve with $\Delta E_Z=|g^{\ast}| \mu_B \Delta B_N^z$ using $(\uparrow,\downarrow)$, $(\downarrow,\uparrow)$, and $(2,0)S$ as a basis set \cite{Supplement}. 
Here, $(\uparrow,\downarrow)$ [$(\downarrow,\uparrow)$] denotes the state with spin-up (-down) and spin-down (-up) electrons in the left and right QDs, respectively. 
The calculation shows that, for $\Delta E_Z>0$, the $(\uparrow,\downarrow)$ component in $S_L^\prime$ becomes progressively dominant with increase in $\Delta E_Z$.
This implies that for $\Delta E_Z>0$ the probability of $T_+\rightarrow S_L^\prime$ transition being mediated by a NS flip in the right QD increases with $\Delta E_Z$. 
The key observation here is that the resultant down-to-up NS flip in the right QD increases $\Delta E_Z$ and further facilitates NS flip in the right QD. 
The same argument holds for $\Delta E_Z<0$, which facilitates NS flips in the left QD. 
Due to this positive feedback, large $\left\vert \Delta B_N^z\right\vert$ can grow spontaneously out of tiny statistical fluctuations. 
This finite $\Delta B_N^z$ produces a dramatic consequence on transport: when $\Delta B_N^z\neq 0$, $T_0^\prime$ has a finite $(2,0)S$ component, which lifts the SB due to $T_0^\prime$. 
With increase in $\left\vert \Delta B_N^z\right\vert$, this $(2,0)S$ component first grows rapidly at small $\left\vert \Delta B_N^z\right\vert$, but then increases only gradually at large $\left\vert \Delta B_N^z\right\vert$, which accounts for the observed sharp onset and subsequent gradual increase of $I_{\text{dot}}$. 
Imbalance in NS polarization has also been invoked to account for the hysteresis and multiple-step behavior of leak current observed in a DQD under ESR condition \cite{Vink,Danon}. 
Despite the marked differences in the experimental situations between the present study and that in Ref.\,\cite{Vink}, $\Delta B_N^z$ plays critical roles in both experiments, which clearly demonstrates that it must be taken into account to fully understand electron-nuclear spin dynamics in DQDs, not only in pulsed-gate, but also in DC-drive experiments.


In summary, we have reported a new hysteretic transport feature in the CB regime of a $\text{GaAs}$ lateral DQD, which signals DNP due to the $T_{+}$-$S_{L}$ crossing. This DNP signal appears owing to $(2,0)S$-hybridization in $S_{L}$ by finite $t$. The magnetic-field dependence of the current level in the DNP region suggests strongly unbalanced NS polarization. 


We thank Y. Okazaki and T. Ota for experimental advice and Y. Tokura and Y. V.
Nazarov for discussions.



\begin{thebibliography}{1}
\bibitem {sequential}This is for the lower left bias-triangle in Fig.\,1(b). For the upper right triangle, $(1, 1) \rightarrow(2, 0) \rightarrow(2, 1) \rightarrow(1, 1)$.
\end{thebibliography}
\end{document}


\makeatletter 
\renewcommand{\figurename}{FIG.}
\renewcommand{\thefigure}{S\arabic{figure}}



\title{Supplementary Material for\\``Observation of Hysteretic Transport Due to Dynamic Nuclear Spin Polarization\\
in a GaAs Lateral Double Quantum Dot''}

\author{Takashi Kobayashi}
\affiliation{NTT\,Basic\,Research\,Laboratories,\,NTT\,Corporation,\,3-1\,Morinosato-Wakamiya,\,Atsugi\,243-0198,\,Japan}
\affiliation{Department of Physics, Tohoku University, Sendai 980-8578, Japan}
\author{Kenichi Hitachi}
\affiliation{NTT\,Basic\,Research\,Laboratories,\,NTT\,Corporation,\,3-1\,Morinosato-Wakamiya,\,Atsugi\,243-0198,\,Japan}
\author{Satoshi Sasaki}
\affiliation{NTT\,Basic\,Research\,Laboratories,\,NTT\,Corporation,\,3-1\,Morinosato-Wakamiya,\,Atsugi\,243-0198,\,Japan}
\affiliation{Department of Physics, Tohoku University, Sendai 980-8578, Japan}
\author{Koji Muraki}
\affiliation{NTT\,Basic\,Research\,Laboratories,\,NTT\,Corporation,\,3-1\,Morinosato-Wakamiya,\,Atsugi\,243-0198,\,Japan}
\date{\today}

\maketitle

\section{Hamiltonian}

\begin{figure}
\includegraphics[width=0.6\linewidth]{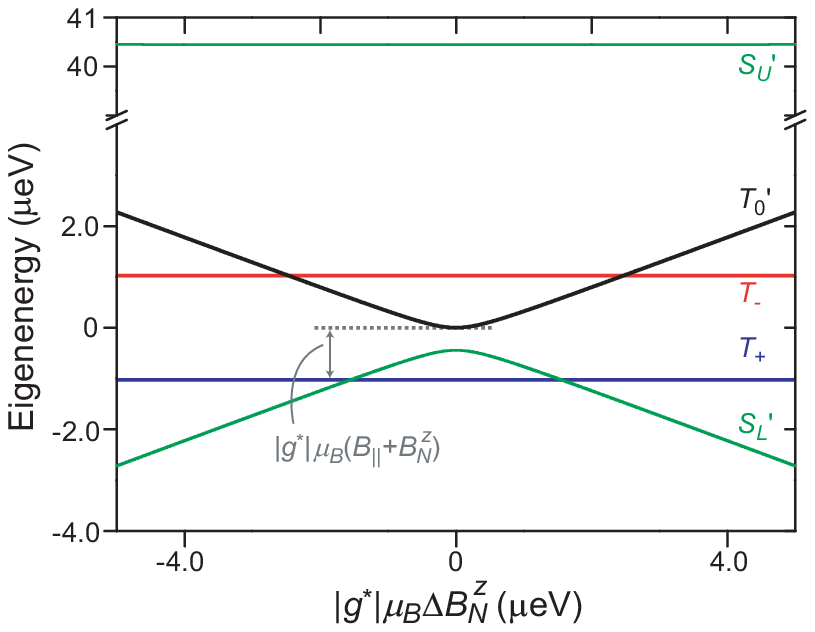}
\caption{\label{fig:figSp1}
Energy level diagram of the five five eigenstates $S_U^\prime, T_0^\prime, S_L^\prime, T_+, T-$ for the $H$ as a function of $g^*\mu_B\Delta B_N^z$. $\varepsilon=-40\,\mu\text{eV}$ and $g^{\ast}\mu_B(B_{\parallel}+B_N^z)=-1.0\,\mu\text{eV}$ are assumed.
}
\end{figure}

Eigen energies are calculated with the Hamiltonian $H=H_{\text{DQD}}+H_N^z$, where $H_{\text{DQD}}$ and $H_N^z$ are
\begin{eqnarray}
H_{\text{DQD}} 
	&=& -\varepsilon |(2, 0)S\rangle \langle (2, 0)S| + \sum_{\sigma=\pm 1}\sigma g^*\mu_{\text{B}}B_{||}|T_\sigma\rangle \langle T_\sigma|\nonumber\\
	& &\ \ \ \ \ + t_C\left[\exp (i\varphi)|(2, 0)S\rangle \langle (\uparrow,\downarrow)| -\exp (i\varphi)|(2, 0)S\rangle \langle (\downarrow,\uparrow)|\right.\nonumber\\
	& &\left. + \exp (-i\varphi)|(\uparrow,\downarrow)\rangle \langle (2, 0)S|-\exp (-i\varphi)|(\downarrow,\uparrow)\rangle \langle (2, 0)S|\right]\nonumber\\
	&=&\left(
	\begin{array}{ccc|cc}
	-\varepsilon & t\exp (i\varphi) & -t\exp (i\varphi) & 0 & 0 \\
	t\exp (-i\varphi) & 0 & 0 & 0 & 0 \\
	-t\exp (-i\varphi) & 0 & 0 & 0 & 0 \\
	\hline
	0 & 0 & 0 & g^*\mu_{\text{B}}B_{||} & 0 \\
	0 & 0 & 0 & 0 & -g^*\mu_{\text{B}}B_{||}
	\end{array}
	\right),
\end{eqnarray}
and
\begin{eqnarray}
H_N^z
	&=& 
	g^*\mu_{\text{B}}\left(
	\Delta B_N^z|(\uparrow,\downarrow)\rangle \langle (\uparrow,\downarrow)|/2
	-\Delta B_N^z|(\downarrow,\uparrow)\rangle \langle (\downarrow,\uparrow)|/2 +B_N^z|T_+\rangle \langle T_+|
	-B_N^z|T_-\rangle \langle T_-|\right) \nonumber\\
	&=&\left(
	\begin{array}{ccc|cc}
	0 & 0 & 0 & 0 & 0 \\
	0 & g^*\mu_{\text{B}}\Delta B_N^z/2 & 0 & 0 & 0 \\
	0 & 0 & -g^*\mu_{\text{B}}\Delta B_N^z/2 & 0 & 0 \\
	\hline
	0 & 0 & 0 & g^*\mu_{\text{B}}B_N^z & 0 \\
	0 & 0 & 0 & 0 & -g^*\mu_{\text{B}}B_N^z
	\end{array}
	\right),
\end{eqnarray}
on a basis set of $(2,0)S$, $(\uparrow,\downarrow)$, $(\downarrow,\uparrow)$, $T_+$, and $T_-$. Note that $|(\uparrow,\downarrow)\rangle =[|T_0\rangle + |(1,1)S\rangle ]/\sqrt{2}$ and $|(\downarrow,\uparrow)\rangle =[|T_0\rangle - |(1,1)S\rangle ]/\sqrt{2}$ for derivation of Figs.\,1(c), 1(d) and 3(f). Figure\,\ref{fig:figSp1} shows typical energy level diagram for the five eigen states as a function of $g^*\mu_B\Delta B_N^z$. Since only the $S_L^\prime$ level can cross at $\varepsilon<0$ with the $T_+$ level from this figure, only the $T_+\rightarrow S_L^\prime$ scattering is considered in our letter.

